\documentclass[conference]{IEEEconf}

\input epsf
\usepackage{graphicx}
\usepackage{multirow}
\usepackage{titlesec}
\usepackage{balance} 
\usepackage{amsmath}
\usepackage{graphicx}
\usepackage{stfloats}
\usepackage{listings}
\usepackage{xcolor,color}
\usepackage{url}
\usepackage[utf8]{inputenc}
\usepackage{indentfirst}
\usepackage{color} 
\usepackage[switch]{lineno}
\usepackage[numbers,sort&compress]{natbib}
\definecolor{codegray}{rgb}{0.5,0.5,0.5}
\definecolor{backcolour}{rgb}{0.95,0.95,0.92}
\definecolor{dkgreen}{rgb}{0,0.6,0}
\definecolor{gray}{rgb}{0.5,0.5,0.5}
\definecolor{mauve}{rgb}{0.58,0,0.82}
\definecolor{preserveword}{rgb}{0,0,255}
\makeatletter
\let\@ORGmakecaption\@makecaption
\long\def\@makecaption#1#2{\@ORGmakecaption{#1}{#2}\vskip\belowcaptionskip\relax}
\makeatother

\renewcommand\thesection{\arabic{section}} 

\renewcommand\thesubsectiondis{\thesection.\arabic{subsection}}

\renewcommand\thesubsubsectiondis{\thesubsectiondis.\arabic{subsubsection}}

\begin{document}

\title{\textbf{\Large An Empirical Study: MEMS as a Static Performance Metric}}

\author{Liwei Zhang$^{1,2,3,4}$, Baoquan Cui$^{2,3}$, Xutong Ma$^{5}$, Jian Zhang$^{1,2,3,4}$\\
	\normalsize $^{1}$Hangzhou Institute for Advanced Study, UCAS, Hangzhou, China\\
	\normalsize $^{2}$Key Laboratory of System Software, CAS, Beijing, China \\
      \normalsize $^{3}$Institute of Software, Chinese Academy of Sciences (CAS), Beijing, China\\
	\normalsize $^{4}$University of Chinese Academy of Sciences (UCAS), Beijing, China\\
	\normalsize $^{5}$Inria, Paris, France\\
	\normalsize \{zhanglw, 
 zj, cuibq\}@ios.ac.cn, ~xutong.ma@inria.fr
}


\maketitle
\begin{abstract}
Static performance estimation is essential during compile-time analysis, yet traditional runtime-based methods are costly and platform-dependent. 
We investigate \textit{mems}, the number of memory accesses, as a static and architecture-independent performance metric.
We develop a Clang-based automated instrumentation tool that rewrites source code to insert path tracing and \textit{mems} counting logic. 
This allows us to evaluate \textit{mems}-based performance estimation across ten classical algorithm programs. Experimental results show that within the same program, execution paths with higher \textit{mems} values consistently exhibit longer runtime. However, this correlation weakens between different programs, suggesting that \textit{mems} is best suited for comparing performance of different execution paths in a program.
\end{abstract}
\IEEEoverridecommandlockouts
\vspace{1.5ex}
\pagestyle{plain}
\tolerance=2000
\setlength{\textfloatsep}{1em}
\setlength{\floatsep}{1em}
%
\IEEEpeerreviewmaketitle
\section{Introduction}


\noindent An important aspect of software quality is performance. Program efficiency is typically characterized by complexity measures such as worst-case or average-case time complexity. However, for real-world programs, even domain experts can find it difficult to accurately predict performance. For example, it took over a decade after the Boyer-Moore string matching algorithm was introduced~\cite{Boyer1977} before its complexity was rigorously established~\cite{Cole1991}. Moreover, theoretical complexity does not always align with practical performance due to hidden constants and implementation-level effects. Thus, empirical benchmarking remains an essential tool for evaluating program behavior.

However, empirical performance evaluation often suffers from limited coverage, environment dependence, and high resource cost. Poor performance not only affects efficiency but also impacts user experience and reliability, making performance analysis a critical part of software quality assurance.

To address the gap between theoretical models and practical behavior, researchers have explored static and symbolic analysis methods. A notable example is Zhang's framework for \textit{performance estimation using symbolic data}~\cite{Zhang2013}, which uses symbolic execution and volume computation~\cite{Ma2009,Ge2021} to estimate performance without running the program. The key idea is to represent execution conditions symbolically and compute the volume of input space satisfying each path, enabling quantitative analysis of potential overheads.

\textbf{Problem.}
Despite these advances, static performance estimation methods remain limited in practicality. Empirical methods require extensive runtime testing and are sensitive to hardware, while symbolic approaches are rarely applied to large-scale C programs and often lack automation. Most symbolic metrics, like operation counts or control-flow complexity, are not validated with real execution data and fail to account for path-level performance variance.

\textbf{Existing Approaches.}
Traditional performance profiling relies on repeated program executions with various inputs to gather timing statistics~\cite{Saumya2019,Bundala2015,Chen2022}. While these techniques are valuable, they may miss rare or worst-case paths and often require significant computational effort. Automated testing frameworks like \textit{XSTRESSOR}~\cite{Saumya2019} and hybrid tools like those from Bundala and Seshia~\cite{Bundala2015} improve test input quality but still struggle with scalability and low-level performance modeling. To improve generalizability and efficiency, symbolic benchmarking~\cite{Zhang2013} emerged—leveraging symbolic execution to explore multiple paths in a single analysis and estimate performance using metrics like comparison count or memory accesses.

\textbf{Our Metric (mems).}
We adopt the \textit{memory access count}—abbreviated as \textbf{mems}—as a machine-independent performance metric. Originally proposed by Knuth~\cite{Knuth1994}, mems represents the number of memory read/write operations in a program and offers a portable estimate of performance. Unlike raw execution time, which is affected by CPU and system environment, mems is a symbolic metric that reflects a program’s inherent memory behavior.

Extending this idea, we design a path-sensitive static analysis that uses symbolic execution to compute memory accesses for each feasible control-flow path. By applying model counting techniques~\cite{Ma2009,Ge2021}, we compute the expected performance as a weighted average over path-level mems. This method enables early-stage and architecture-neutral performance reasoning without requiring actual execution.

\textbf{Challenges.}
Although mems has been proposed in theoretical discussions~\cite{Zhang2013, Knuth1994}, no prior work has systematically evaluated its practicality in real software analysis. Key challenges include: (1) automatically extracting memory access counts from C programs; (2) validating mems against real-world performance metrics like execution time; and (3) building scalable tools for large codebases and numerous paths.

\textbf{Solutions.}
To address these challenges, we propose and implement the following:
\begin{itemize}
\item \textbf{New Metric.} We define \textit{mems} as a lightweight, architecture-independent performance metric based on memory access counts, emphasizing array operations as major contributors to execution cost.

\item \textbf{Automated Solution.} We develop an automated pipeline for mems analysis based on the \texttt{eppather} tool~\cite{eppather}, derived from the \texttt{epat} test generation framework~\cite{XuZhang2006,JZhang2004}. Our tool integrates AST-based analysis, path traversal, and source-to-source Clang instrumentation to collect mems and timing data.

\item \textbf{Evaluation.} We perform extensive experiments to evaluate whether mems correlates with execution time across (1) different paths within the same program (RQ1), and (2) across different programs and environments (RQ2). Results on multiple platforms demonstrate that mems effectively approximates performance in many cases, highlighting its utility and limitations.
\end{itemize}

The findings of this empirical research contribute directly to improving the accuracy and reliability of static performance indicators used in symbolic benchmarking. By clarifying the strength and limitations of \textit{mems} as a performance predictor, this study does not only enhance the understanding of program complexity but also provides practical guidance for more effective performance evaluation, optimization strategies, and software testing practices.

\section{Preliminary}

\noindent Static program analysis plays a vital role in program understanding, optimization, and verification. One commonly used representation is the control flow graph (CFG), where nodes represent statements or basic blocks, and edges represent possible control flow between them. Many static analysis techniques, such as abstract interpretation and symbolic execution, operate on CFGs to detect bugs, verify properties, or estimate performance.

\subsection{The {\it mems\/} Metric}

\noindent Usually we evaluate a program's performance by executing it on a computer with some input data,
and measure its duration of execution. This can be done repeatedly with different input data.
As abstracted as modeled with indicated in Moore's Law, the improvement in computer devices leads to an increment in the efficiency of program execution as time passes. For many years, computers have been getting faster and faster.
Thus a program's running times (in seconds) are getting less and less.

Knuth came up with an idea to measure a program's performance, which is independent of the processor's computing power.
He proposed to use the number of memory accesses (in short, {\it mems\/}) ~\cite{Knuth1994}to indicate the execution speed of a program.Knuth has evaluated various SAT solvers using the mems metric ~\cite{Knuth2012}.
We simply introduce a special variable -- {\tt mems}, which is initialized to 0 and incremented every time a memory access (such as an array read or write) occurs during execution.
In a statement which involves memory access, we add a statement to increase the value of {\tt mems}.
Knuth often uses the notation $o$ to increase the value by 1, $oo$ to increase the value by 2, and so on.
Thus we may have statements like the following, to compute the value of {\tt mems}.
\begin{verbatim}
        o, arr[i] = i+2;
       oo, arr[i+1] = arr[i];
\end{verbatim}

The {\it mems\/} has not been used widely. In the subsequent sections of this paper, we will conduct a series of experiments to check whether this indicator is feasible.
\subsection{Path-Based Performance Estimation}

\noindent A widely adopted method in static analysis is \textit{path-sensitive analysis}, where a program is analyzed along individual execution paths extracted from the CFG ( Control Flow Graph ). Each path begins at the program’s entry and ends at an exit point. Analyzing individual paths allows for fine-grained understanding of the program’s behavior.

However, not every path generated from the CFG is executable. To determine if a path is feasible, \textit{symbolic execution} is employed to derive the \textit{path condition}, which is a conjunction of constraints over program inputs. A path is considered feasible only if its path condition is satisfiable. SMT ( Satisfiability Modulo Theories ) solvers can check the satisfiability~\cite{Ma2009,Zhang2008,Zhang2013draft}, and model counters can compute how many input assignments satisfy the path condition, which we refer to as the \textit{path frequency}.

To estimate performance in a machine-independent manner, we define \textit{Performance Indicators} (PINDs), such as the number of memory accesses, arithmetic operations, or comparisons on a given path. Let a program consist of a set of paths $\{P_i\}$. Each path $P_i$ has~\cite{Zhang2004}:
\begin{itemize}
  \item A frequency $\delta_i$ — how often the path is executed (based on the number of satisfying inputs).
  \item A PIND value $\text{pind}_i$ — the cost incurred along the path (e.g., number of memory operations).
\end{itemize}

The estimated performance of the entire program is then computed as the weighted average:
\[
\text{Performance} = \frac{\sum_i (\delta_i * \text{pind}_i)}{\sum_i \delta_i}
\]

This technique has been extended to characterize not only average cost but the full distribution of performance across the input space. For example, Chen et al.~\cite{Chen2016} proposed using probabilistic symbolic execution to derive \textit{performance distributions} rather than single-point estimates. This enables modeling not just expected runtime, but also the variance and tail behaviors of performance under uncertainty.

To represent conditional branches along an execution path, we use the notation \texttt{@ (condition)} to denote the occurrence of a branching condition within the code. This can be done via symbolic execution~\cite{King1976}.

\subsubsection{Example} 

Consider the following code snippet:

\begin{lstlisting}
int x;
@ ( (x > 20) && (x <= 100) )
x = x - 10; if (x > 30) {...} else {...} 
\end{lstlisting}

This format is also used in our instrumentation output to log the sequence of conditions encountered during execution.
This program has two feasible paths:
\begin{itemize}
  \item $P_T$: takes the branch where \texttt{x - 10 > 30}
  \item $P_F$: takes the branch where \texttt{x - 10 <= 30}
\end{itemize}

The corresponding path conditions are:
\begin{lstlisting}
(*|$P_T$|*): (x > 20) && (x <= 100) && (x - 10 > 30)
(*|$P_F$|*): (x > 20) && (x <= 100) && (x - 10 <= 30)
\end{lstlisting}

With a model counting tool~\cite{Geldenhuys2012,Liu2011,Ge2021}, we will have: $\delta_T = 60$ and $\delta_F = 20$.
For simplicity, we assume the following \texttt{mems} values for the two branches as hypothetical examples: $\text{pind}_T = 3$ and $\text{pind}_F = 2$.
Then the overall performance estimate is:
\[
\text{Performance} = \frac{60 * 3 + 20 * 2}{60 + 20} = \frac{260}{80} = 2.75
\]

This value reflects the average number of operations to be executed with any concrete inputs in the input domain.

\subsubsection{Remarks} 

This method is particularly useful in early-stage design and static optimization, where empirical timing information is not yet available. It can also guide compiler decisions or resource estimation in embedded systems, where performance metrics like memory operations are more meaningful than raw execution time.

\section{Approach}

\noindent The concept of using memory access counts, \textit{mems}, as a machine-independent performance indicator was first proposed by Knuth. In his book \textit{The Stanford GraphBase}~\cite{Knuth1994} and further discussed in the \textit{The Art of Computer Programming}~\cite{Knuth1997}, Knuth introduced the idea of instrumenting programs with a \texttt{mems} counter to reflect the computational cost associated with memory operations, independent of processor speed.

Building on this idea, our previous work~\cite{Zhang2013} proposed a symbolic estimation framework in which \texttt{mems} is calculated along individual control-flow paths. The paper further suggested aggregating these per-path \textit{mems} values using weighted averages based on symbolic path frequencies to obtain a performance estimate for the entire program.

However, despite the elegance of this theory, none of these studies provided empirical validation on real-world programs. As a result, it remains unclear whether \texttt{mems} values meaningfully correlate with execution time in practice, especially for modern C programs across various input paths and platforms.

To bridge this gap, we perform large-scale experiments by executing instrumented C programs and collecting path-level metrics—\texttt{mems}, path length, and runtime. This requires the construction of a precise and automated instrumentation framework, detailed below.

\subsection{Instrumentation Pipeline}

\noindent To facilitate data collection, we built a Clang-based tool named \texttt{eppather-clangpass}. This tool integrates directly with Clang’s AST infrastructure and modifies the source code by inserting instrumentation logic into user-defined functions (excluding \texttt{main}). During execution, the instrumented program records the control-flow path, memory access count (\texttt{mems}), path length, and timing metrics, all printed to standard output for batch analysis.
The instrumentation pipeline operates in the following stages:

\begin{itemize}
  \item \textbf{Function Entry and Exit.} At the entry of each user-defined function, we insert definitions and initializations for the variables used to track performance metrics, including \texttt{mems}, \texttt{path\_len}, and timing variables such as \texttt{start}, \texttt{end}, and \texttt{freq} (on Windows, using \texttt{QueryPerformanceCounter} APIs). At the function exit (before every \texttt{return} statement), we insert logic to compute and print the total execution time, memory access count, and path length.

  \item \textbf{Conditional and Loop Statements.} For every conditional or loop predicate (e.g., \texttt{if}, \texttt{while}, \texttt{for}), we inject a \texttt{printf} statement to record the evaluation of the branch condition. The true branch is annotated as \texttt{@(condition)}, while the false branch is annotated as \texttt{@(!(condition))}, where \texttt{condition} is the string representation of the original source predicate. Each branch point also increments the path length counter.

  \item \textbf{Assignment Statements.} Each assignment operation is instrumented with a \texttt{printf} that logs the source-level statement as a step in the execution path. This captures the concrete path taken through the program logic.

  \item \textbf{Array Access Detection.} For assignments or expressions involving array reads or writes (e.g., \texttt{arr[i] = ...} or \texttt{... = arr[i]}), we increment the \texttt{mems} counter accordingly. For example, a statement like \texttt{arr[i] = arr[i] + 1} is counted as two memory operations— read and write.

\end{itemize}

This instrumentation strategy allows us to track symbolic path conditions, memory usage, and runtime characteristics in a unified and automated way. The resulting output can then be parsed for further batch analysis to assess the relationship between static metrics (like \texttt{mems}) and dynamic behavior (like execution time).

\subsection{Instrumentation Example}

\noindent The following listing shows a simplified instrumented version of a loop-based test program used in RQ1. 
\textit{Red-highlighted lines} represent the original logic of the program (e.g., control structures and memory operations), 
while the \textit{non-highlighted lines} correspond to the inserted instrumentation code. 
Instrumentation logic is placed before and after control branches and memory-related expressions to ensure precise path tracking, 
memory access counting, and execution time measurement.

\begin{figure}[!ht]
\centering
\begin{lstlisting}[caption={Simplified instrumented version with original code highlighted}, label={lst:highlighted}]
LARGE_INTEGER freq, start, end;
QueryPerformanceFrequency(&freq);
QueryPerformanceCounter(&start);
(*|\textcolor{red}{int a = 0, b = 0, i;}|*)
printf("Path:\n");
(*|\textcolor{red}{for (i = 0; i < n; i++)}|*) {
    printf("#(i < n)\n"); 
    path_len = path_len + 1;
    (*|\textcolor{red}{if (mode > 0)}|*) {
        printf("#(mode > 0)\n"); 
        path_len = path_len + 1;
        (*|\textcolor{red}{arr[i] = i * 2 + arr[i];}|*) 
        mems = mems + 2;
        (*|\textcolor{red}{mode = mode - 1;}|*)
    } (*|\textcolor{red}{else }|*){
        printf("#(!(mode > 0))\n"); 
        path_len = path_len + 1;
        (*|\textcolor{red}{b = i * 3 + b;}|*)
        (*|\textcolor{red}{mode = mode - 1;}|*)
    }
}
printf("Total path length: %d\n", path_len);
printf("Total memory accesses: %d\n", mems);
QueryPerformanceCounter(&end);
double time_taken = (double)(end.QuadPart - start.QuadPart) / freq.QuadPart;
printf("Execution time: %f\n", time_taken);
\end{lstlisting}
\end{figure}

This instrumented version tracks control-flow conditions using annotated \texttt{printf} statements, counts array memory accesses using the \texttt{mems} counter, and measures execution time using high-precision timers.






\section{Evaluation}
\noindent We evaluate the practicality and predictive power of using memory access counts (\texttt{mems}) as a static performance metric through empirical experiments guided by the two research questions outlined above. Each question is addressed via multiple targeted validations as follows:

\bigskip

\noindent\textbf{RQ1: Intra-Program Correlation Between \texttt{mems} and Execution Time.}
This part of the evaluation aims to determine whether \texttt{mems} values can effectively indicate performance differences among paths within the same program.

\begin{itemize} \item \textbf{Validation 1: Equal-Length Paths.}
For paths of equal length within the same program (i.e., same number of control-flow decisions), we examine whether higher \texttt{mems} values consistently result in longer execution times. A motivating example program is constructed with a loop and branching logic where some paths involve memory accesses while others do not.

\item \textbf{Validation 2: Path-Wide Correlation.}
We collect data for all feasible execution paths in a program and compute statistical correlations (e.g., Pearson correlation coefficient) between \texttt{mems} values and corresponding execution times. This evaluates whether \texttt{mems} provides a monotonic signal for performance under realistic control-flow variations. \end{itemize}

\bigskip

\noindent\textbf{RQ2: Cross-Program Generalizability of \texttt{mems} as a Performance Metric.}
This part investigates whether \texttt{mems} remains a useful indicator of performance across multiple programs and environments.

\begin{itemize} \item \textbf{Validation 1: Server-Side Analysis.}
We execute a set of classic algorithm implementations on a high-performance Linux server, measuring runtime using both the system \texttt{time} command and Valgrind profiling. We compare \texttt{mems} values with observed execution times across different programs.

\item \textbf{Validation 2: Optimization-Free Local Execution.}
To reduce noise from compiler optimizations, we rerun all tests on a local Windows machine with optimization flags disabled (\texttt{-O0}). Execution time is recorded using high-precision timing APIs, providing an additional perspective on the \texttt{mems}-time relationship.

\item \textbf{Validation 3: Single-Core Execution.}
We further isolate CPU-level interference by pinning execution to a single core, thus minimizing variability from thread scheduling and multi-core interactions. This allows us to assess whether \texttt{mems} remains predictive under strict serial execution. \end{itemize}

These experiments collectively examine both path-level and program-level implications of \texttt{mems}, helping assess when and where it offers meaningful insight into performance.

\subsection{Experimental Setup}

To evaluate the effectiveness of \texttt{mems} as a static performance metric, we conducted empirical studies across two research questions (RQ1 and RQ2), spanning multiple test environments and benchmark programs. This section summarizes the program corpus, hardware configurations, and timing methodologies used throughout the experiments.

\subsubsection{Program Benchmarks}

All experiments, unless otherwise stated, were performed on a suite of ten classical C benchmark programs, including sorting, searching, and array-manipulating algorithms. These programs were selected for their control-flow complexity and frequent memory access behavior. For RQ1’s first validation only, we used a specially crafted illustrative program (\texttt{example.c}) designed to isolate the impact of memory operations under constant path lengths.

\subsubsection{Execution Environments}

The experiments were carried out on two platforms:

\begin{itemize}
    \item \textbf{High-Performance Server (RQ1-V1, RQ1–V2, RQ2–V1)}:
    \begin{itemize}
        \item OS: Ubuntu 20.04.6 LTS (x86\_64), Kernel: 5.4.0-204-generic
        \item CPU: Multi-core server (high concurrency), avg. system load: \textasciitilde30.8
        \item Memory: 225~GB RAM (180~GB available)
        \item Timing Method: \texttt{valgrind} (\texttt{callgrind}) used to collect amplified, hardware-agnostic execution time
    \end{itemize}

    \item \textbf{Local Machine (RQ2–V2, RQ2–V3)}:
    \begin{itemize}
        \item OS: Windows 11 Home Edition
        \item CPU: AMD Ryzen 9 7945HX, 2.50 GHz
        \item Memory: 16~GB RAM
        \item Compiler: MSVC with \texttt{-O0} optimization disabled
        \item Timing Method: \texttt{QueryPerformanceCounter} used for high-resolution wall-clock time measurement
        \item RQ2–V3 additionally fixed CPU affinity to enforce single-core execution
    \end{itemize}
\end{itemize}

\subsubsection{Measurement Protocol}

In all configurations, we systematically varied at least one input parameter: program size (denoted as \texttt{n}) to influence both path length and memory access frequency. For each configuration:

\begin{itemize}
    \item Each program was run 5 times per path, and the average execution time was recorded.
    \item Valgrind’s \texttt{callgrind} was used to obtain stable, instruction-level time metrics, especially under high server load.
\end{itemize}

\subsection{RQ1: More Memory Accesses, Longer Execution Time}

\noindent To examine the relationship between memory access counts and execution time under controlled conditions, we first study a single, well-designed C function. This function is crafted to isolate the effect of memory accesses while keeping the path length constant, making it ideal for verifying whether increased \texttt{mems} alone can influence execution time within the same program.

\subsubsection{A Test Case: Memory Accesses Impact Execution Time}
\noindent We begin with a simple C function designed with a loop containing a conditional branch, as shown in Listing~\ref{lst:code}. In this function, the true branch modifies an array (incurring memory accesses), while the false branch performs only arithmetic operations. This setup offers a controlled environment to analyze the impact of memory accesses on execution time, while keeping the path length consistent across different loops.
\begin{figure}[t]
\centering
\begin{lstlisting}[language=C, caption={Function with conditional memory access}, label=lst:code]
void test(int n, int mode) {
    int arr[n], a = 0, b = 0;
    for(int i = 0; i < n; i++) {
        if(mode > 0) {
            arr[i] = i * 2 + arr[i];
            mode = mode - 1;
        } else {
            b = i * 3 + b;
            mode = mode - 1;
        }
    }
}
\end{lstlisting}
\end{figure}

Table \ref{tab:results} presents the refined experimental results collected from uninstrumented programs. Each row corresponds to a specific execution path controlled by two input parameters:

\begin{itemize}
  \item \textbf{p$_1(n)$}: The first parameter, representing the input size of the program, corresponding to the variable \texttt{n} in the snippet.
  \item \textbf{p$_2$}: The second parameter, denoted as \texttt{mode} in the program, controls whether the execution path takes the memory-intensive (true) branch or the lightweight (false) branch during loop iterations.
\end{itemize}

The remaining columns report various metrics collected during execution:
\begin{itemize}
  \item \textbf{t$_0$(ms)}: The actual execution time (in milliseconds) of the original, uninstrumented program.
  \item \textbf{vg$_0$(ms)}: The execution time (in milliseconds) as reported by \texttt{valgrind}'s \texttt{callgrind} tool, providing a hardware-agnostic proxy for instruction cost.
\end{itemize}

These metrics provide insight into how the number of memory operations correlates with execution time under different program scales and path conditions.
For higher values of $n$, where memory operations become significant, a consistent proportional relationship was observed between increased mems and longer execution times, supporting the hypothesis that mems can be a significant indicator of execution time in similar path lengths.

\begin{table}[t]
\centering
\caption{Refined Experimental Results with Execution Time and Valgrind Time on Uninstrumented Programs}
\label{tab:results}
\setlength{\tabcolsep}{2pt}
\begin{tabular}{r|r|c|r|r|c|r}
\hline
\textbf{p$_1(n)$} & \textbf{p$_2$~~} & \textbf{file} & \textbf{len~~} & \textbf{mems~} & \textbf{t$_0$(ms)} & \textbf{vg$_0$(ms)} \\
\hline
10 & 0 & path\_0 & 20 & 0 & 0.031 & 6.538 \\
10 & 5 & path\_1 & 20 & 10 & 0.026 & 6.706 \\
10 & 10 & path\_2 & 20 & 20 & 0.030 & 6.573 \\
50 & 0 & path\_3 & 100 & 0 & 0.038 & 6.762 \\
50 & 25 & path\_4 & 100 & 50 & 0.033 & 7.099 \\
50 & 50 & path\_5 & 100 & 100 & 0.035 & 6.753 \\
100 & 0 & path\_6 & 200 & 0 & 0.068 & 8.488 \\
100 & 50 & path\_7 & 200 & 100 & 0.061 & 8.189 \\
100 & 100 & path\_8 & 200 & 200 & 0.052 & 7.762 \\
500 & 0 & path\_9 & 1000 & 0 & 0.241 & 12.818 \\
500 & 250 & path\_10 & 1000 & 500 & 0.178 & 12.289 \\
500 & 500 & path\_11 & 1000 & 1000 & 0.187 & 12.399 \\
10000 & 0 & path\_30 & 20000 & 0 & 0.079 & 5.729 \\
10000 & 2500 & path\_31 & 20000 & 2500 & 0.092 & 5.555 \\
10000 & 5000 & path\_32 & 20000 & 5000 & 0.089 & 5.838 \\
10000 & 7500 & path\_33 & 20000 & 7500 & 0.116 & 5.718 \\
10000 & 10000 & path\_34 & 20000 & 10000 & 0.128 & 5.538 \\
50000 & 0 & path\_35 & 100000 & 0 & 0.362 & 8.015 \\
50000 & 12500 & path\_36 & 100000 & 12500 & 0.418 & 9.257 \\
50000 & 25000 & path\_37 & 100000 & 25000 & 0.546 & 8.718 \\
50000 & 37500 & path\_38 & 100000 & 37500 & 0.630 & 9.018 \\
50000 & 50000 & path\_39 & 100000 & 50000 & 0.639 & 8.824 \\
100000 & 0 & path\_40 & 200000 & 0 & 0.720 & 9.824 \\
100000 & 25000 & path\_41 & 200000 & 25000 & 0.857 & 10.585 \\
100000 & 50000 & path\_42 & 200000 & 50000 & 1.051 & 11.500 \\
100000 & 75000 & path\_43 & 200000 & 75000 & 1.201 & 12.427 \\
100000 & 100000 & path\_44 & 200000 & 100000 & 1.361 & 14.158 \\
200000 & 0 & path\_45 & 400000 & 0 & 1.319 & 15.259 \\
200000 & 50000 & path\_46 & 400000 & 50000 & 1.662 & 16.742 \\
200000 & 100000 & path\_47 & 400000 & 100000 & 2.077 & 18.213 \\
200000 & 150000 & path\_48 & 400000 & 150000 & 2.467 & 19.587 \\
200000 & 200000 & path\_49 & 400000 & 200000 & 2.544 & 21.368 \\
\hline
\end{tabular}
\end{table}

\begin{figure*}[htbp]
\centering
\begin{minipage}{0.49\textwidth}
    \centering
    \includegraphics[width=\textwidth]{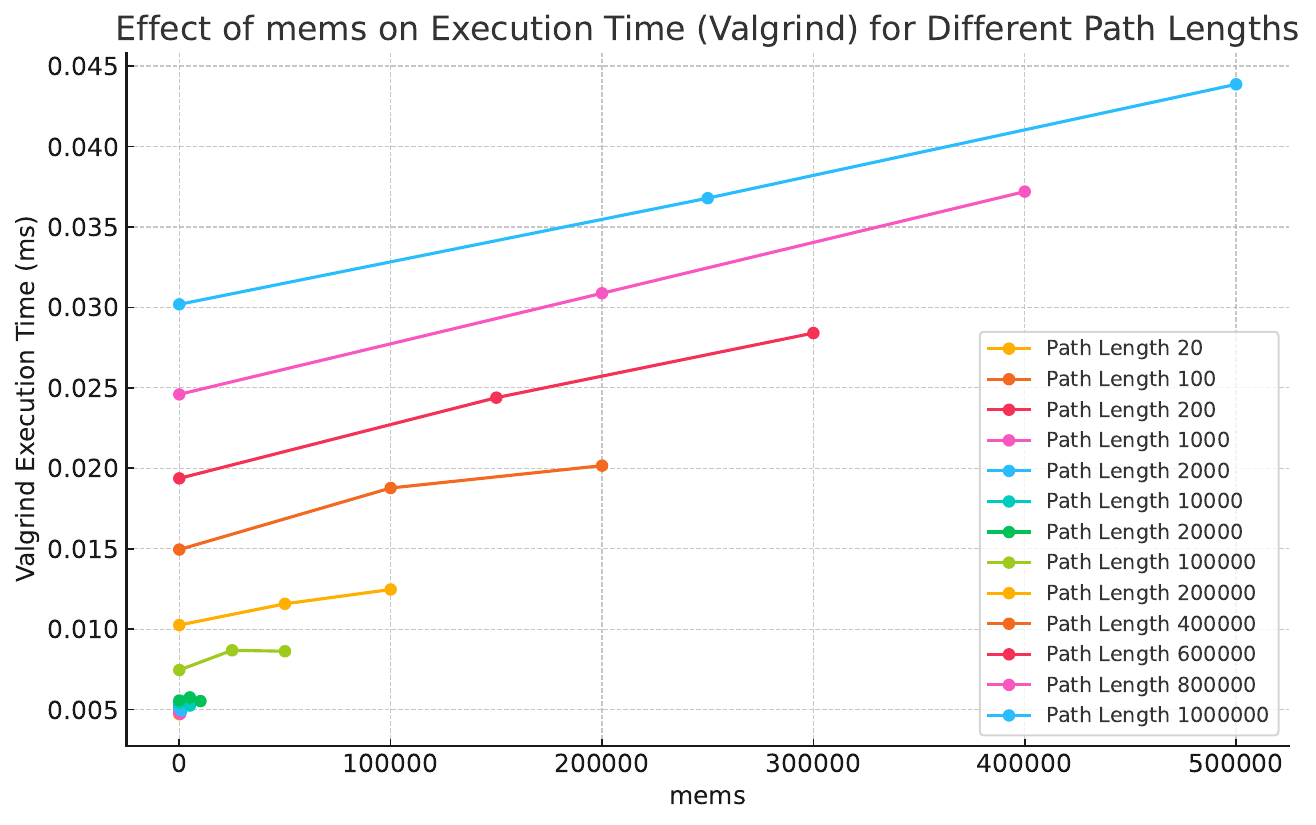}
    \caption{Graphical representation of memory access frequency and execution time correlation.}
    \label{fig:example1}
\end{minipage}\hfill
\begin{minipage}{0.49\textwidth}
    \centering
    \includegraphics[width=\textwidth]{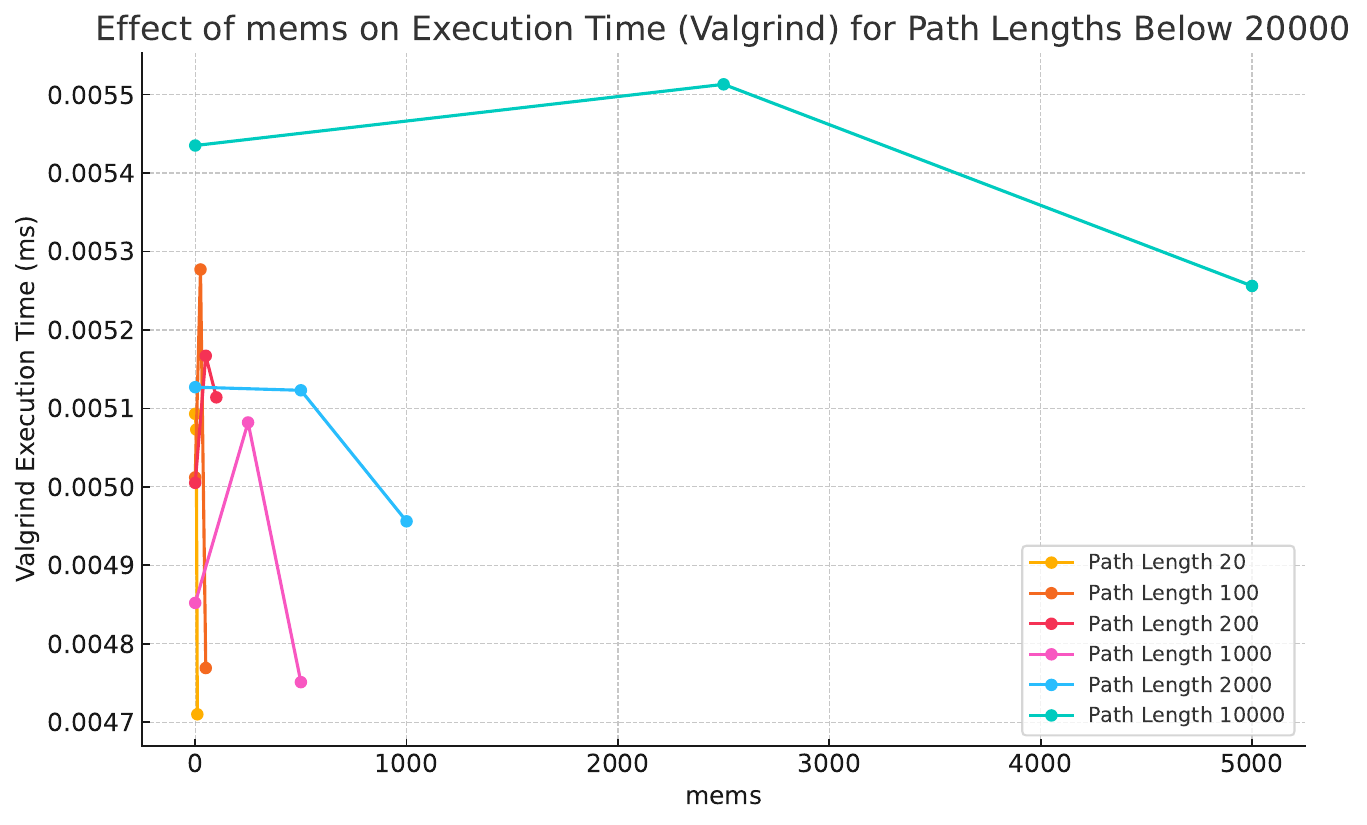}
    \caption{Analysis of path length impact on execution time across different test conditions.}
    \label{fig:example2}
\end{minipage}
\end{figure*}

Figures~\ref{fig:example1} and \ref{fig:example2} illustrate the relationship between \texttt{mems} and Valgrind execution time under fixed path lengths. When the path length is relatively large (e.g., greater than 10,000), a clear positive correlation emerges: higher \texttt{mems} values generally correspond to longer execution times. However, for shorter paths (e.g., path length less than 1,000), this trend becomes less consistent. The execution times in these small-scale cases are more susceptible to measurement noise and system-level fluctuations, which obscure the underlying correlation and reduce the predictive utility of \texttt{mems} in such contexts.

These findings validate the proposed theory and reinforce the importance of considering memory accesses as a key factor in performance analysis, especially in scenarios with similar computational paths.

\subsubsection{Mems–Time Correlation Across Different Path Lengths}

\noindent While the single test case in RQ1 demonstrated a strong correlation between memory accesses and execution time within a single program, it remains unclear whether this relationship holds across different programs with varying control structures and memory usage patterns. To investigate the generality and limitations of \texttt{mems} as a performance predictor, we conducted a comparative study involving multiple benchmark programs. This section presents the experimental setup and findings from that cross-program analysis.

\subsubsection{Results and Observations}
The correlation coefficients were calculated to determine the strength of the relationship between memory usage and execution time across different program paths. The results are presented in Table \ref{tab:correlation_results}.
\begin{table}[t]
\centering
\caption{Correlation Coefficients between Memory Usage and Execution Time}
\label{tab:correlation_results}
{\small 
\begin{tabular}{l|c|c}
\hline
\textbf{Program} & \textbf{Correlation Coefficient} & \textbf{Interpretation} \\
\hline
Array & 0.9998~ & Very strong \\
Bubble & 0.9998~ & Very strong \\
Insertsort & 0.99996 & Very strong \\
Sieve & 0.99986 & Very strong\\
Topo & 0.9990~ & Very strong\\
\hline
\end{tabular}
}
\end{table}

The results indicate a high degree of correlation between \texttt{mems} and execution time for the algorithms assessed within the experimental environment. This observation aligns with the results of specially constructed test cases, supporting the initial hypothesis of RQ1. The findings preliminarily validate the use of \texttt{mems} as a static indicator capable of reflecting performance costs across different paths, thereby substantiating the conjecture posed in RQ1.

\textbf{Answer to RQ1 (Finding 1).} These results demonstrate the feasibility of using \texttt{mems} as a performance metric to analyze the efficiency of various algorithmic paths. The strong correlations observed suggest that memory usage can reliably predict execution time, making it a valuable tool for optimizing algorithm performance in similar settings.

\subsection{RQ2: Mems–Time Correlation Across Programs}
\noindent The motivating example previously presented indicates that, given identical program path lengths, the size of memory accesses (mems) impacts execution time significantly. Following Knuth's theoretical framework, it is expected that programs with similar path lengths and mems across different codes exhibit similar execution times. To examine whether mems can be a reliable performance indicator across diverse codebases, we conducted experiments using ten benchmark programs focused primarily on array-based operations such as sorting and searching algorithms.

\subsubsection{Server-Side Evaluation: Partial Correlation but Inconsistent Predictability}
\noindent Experiments on the server utilized the same environment as described in RQ1. Execution times measured via \texttt{valgrind} correlated strongly (correlation coefficient of 0.98) with those obtained using the Linux \texttt{time} command, validating \texttt{valgrind}'s measurements as suitable for further analysis.

\begin{figure*}[htbp]
    \centering
    \includegraphics[width=0.48\textwidth]{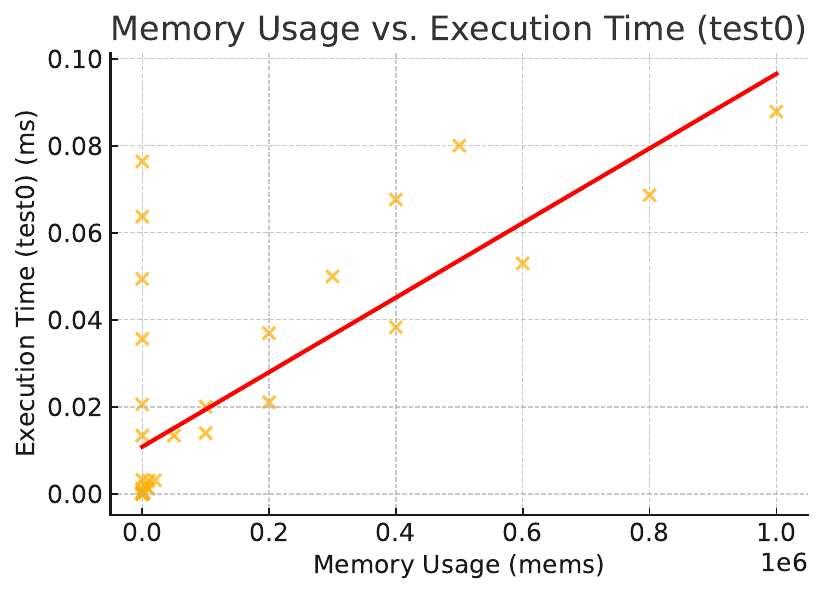}
    \includegraphics[width=0.48\textwidth]{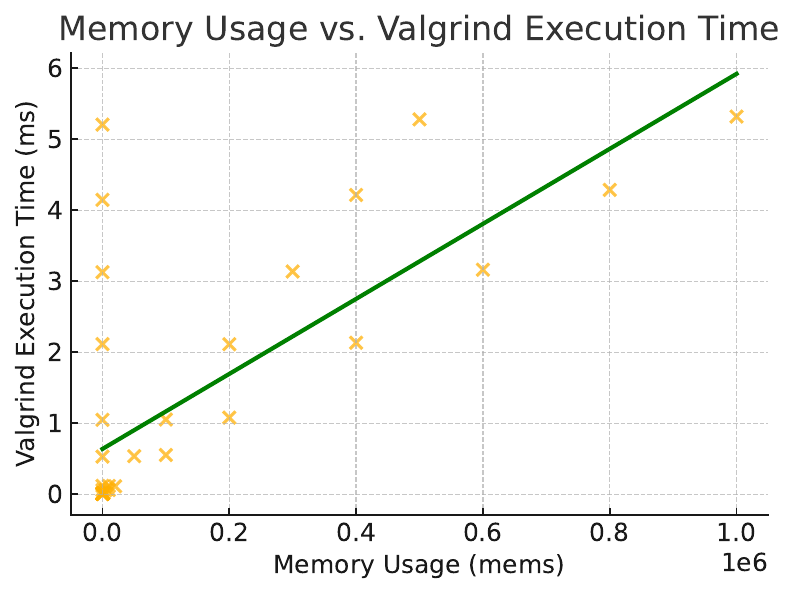}\\
    \includegraphics[width=0.48\textwidth]{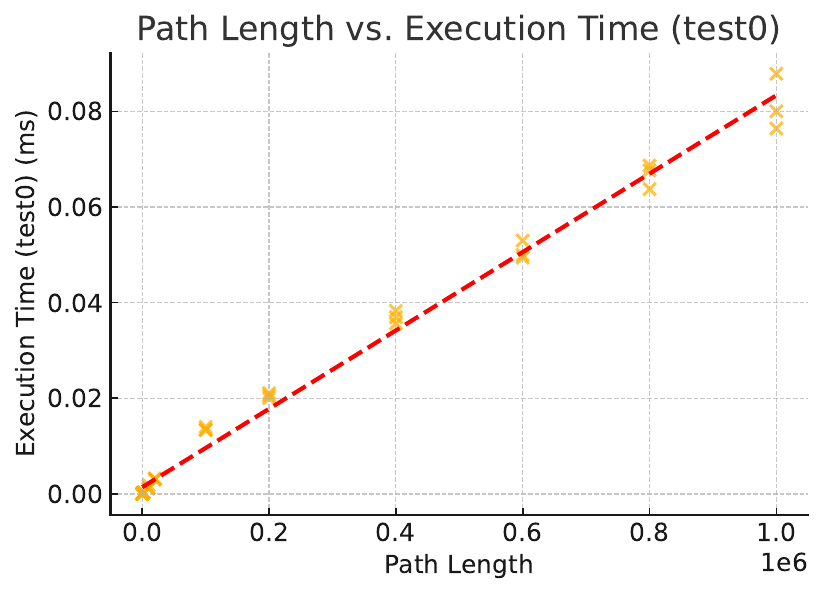}
    \includegraphics[width=0.48\textwidth]{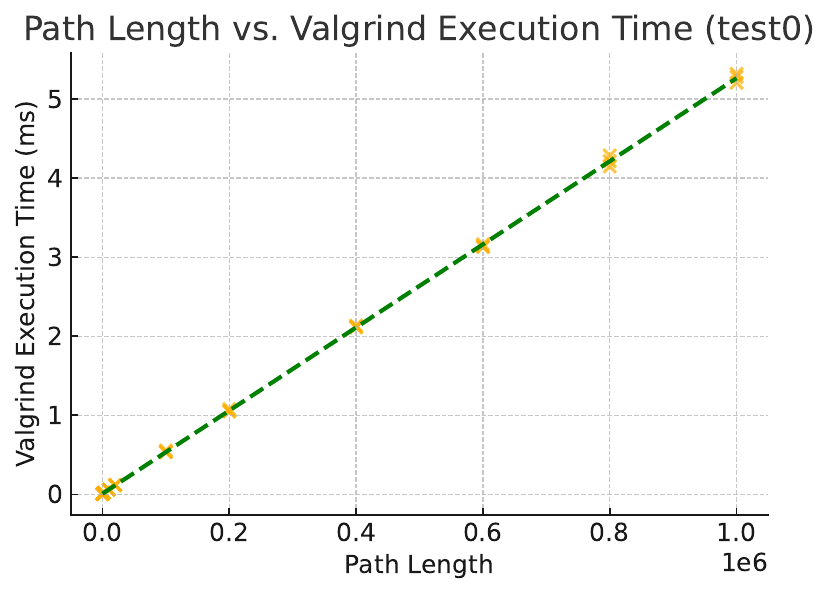}
    \caption{Scatter plots illustrating the relationship between \texttt{mems}, \texttt{path\_length}, and execution time across different experimental scenarios.}
    \label{fig:mems_pathlength_exec_time}
\end{figure*}

As shown in Fig.~\ref{fig:mems_pathlength_exec_time}, the combined scatter plots clearly illustrate general positive correlations between \texttt{mems}, \texttt{path length}, and execution time across different experimental conditions. Despite this general trend, substantial deviations and outliers remain, suggesting that neither \texttt{mems} nor \texttt{path length} individually provides a fully consistent predictor of execution time.

\begin{figure*}[htbp]
    \centering
    \includegraphics[width=0.48\textwidth]{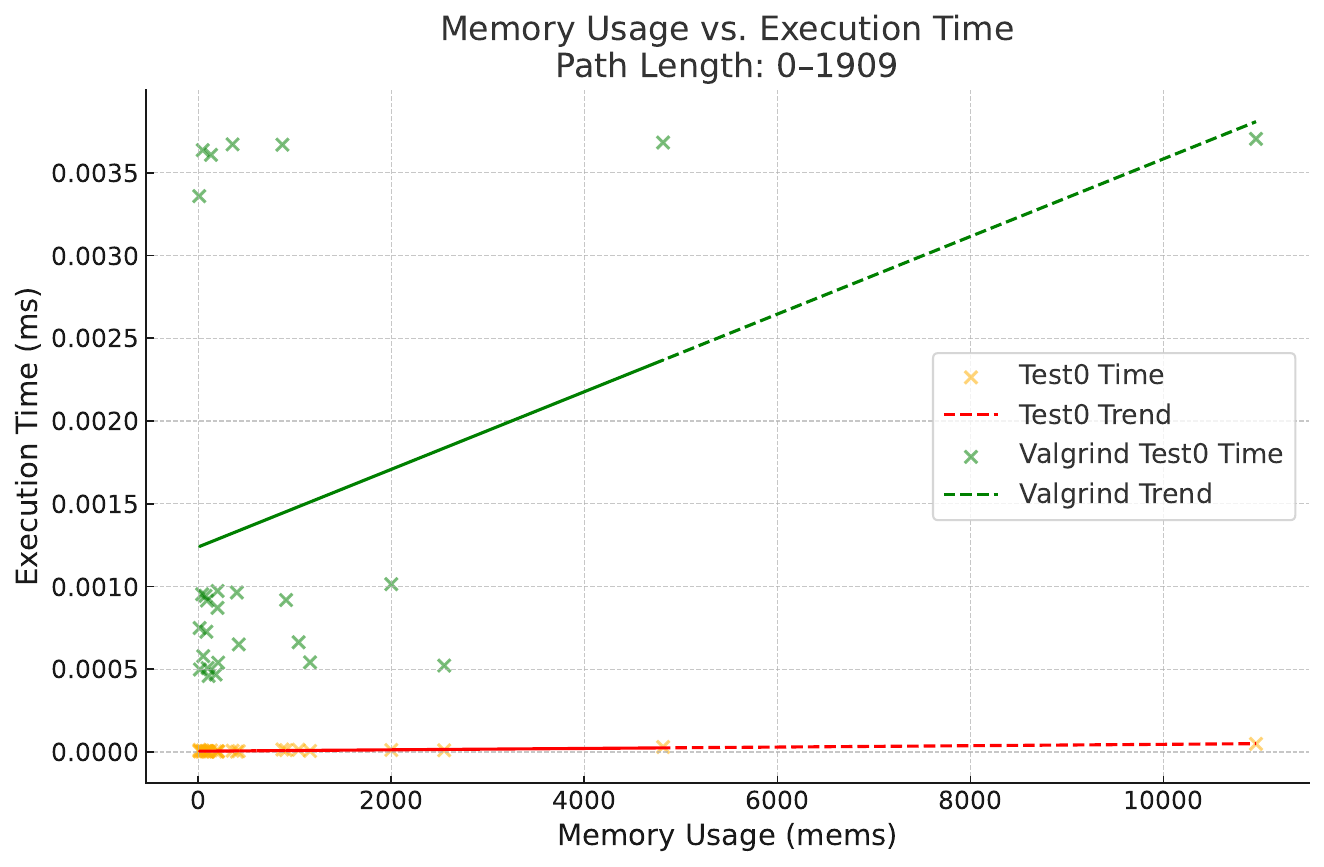}
    \includegraphics[width=0.48\textwidth]{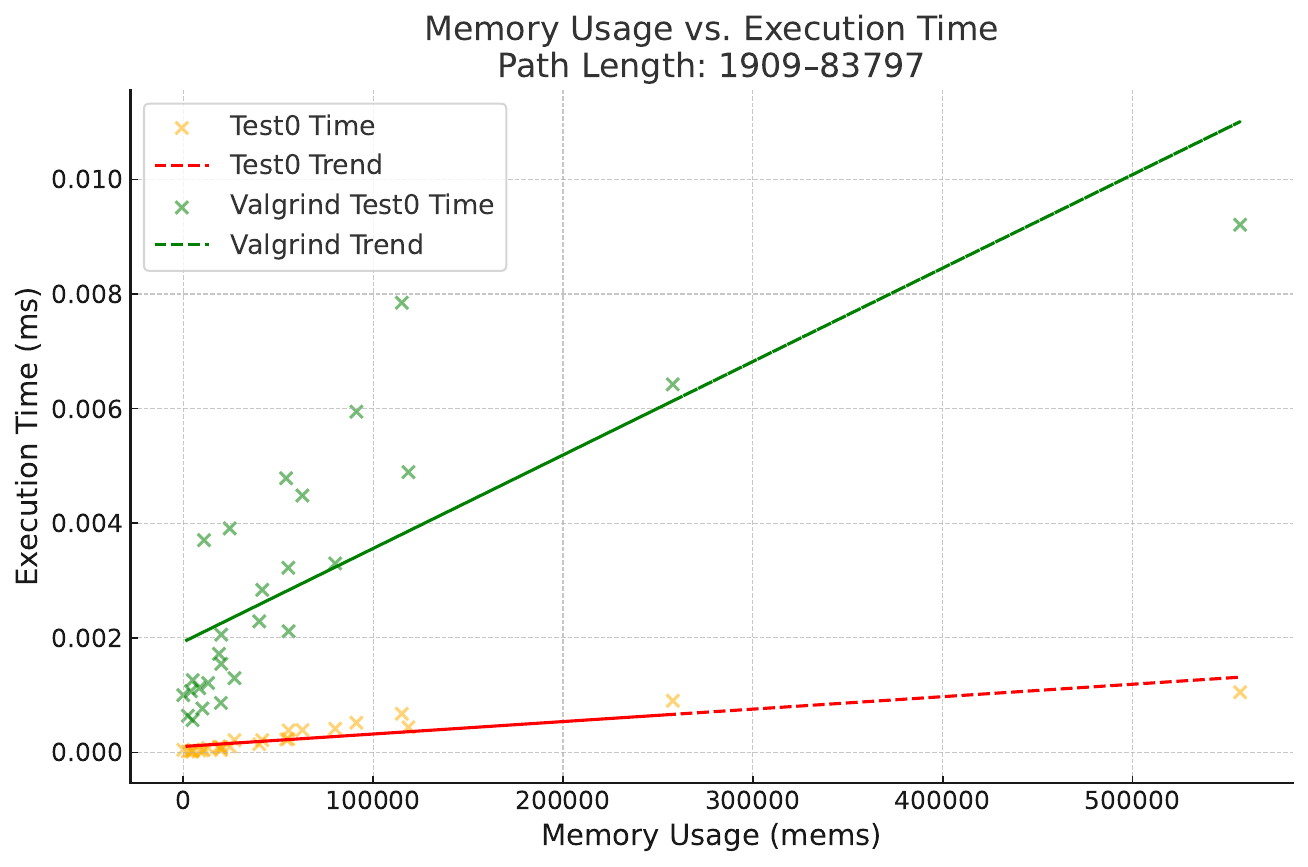}\\
    \includegraphics[width=0.48\textwidth]{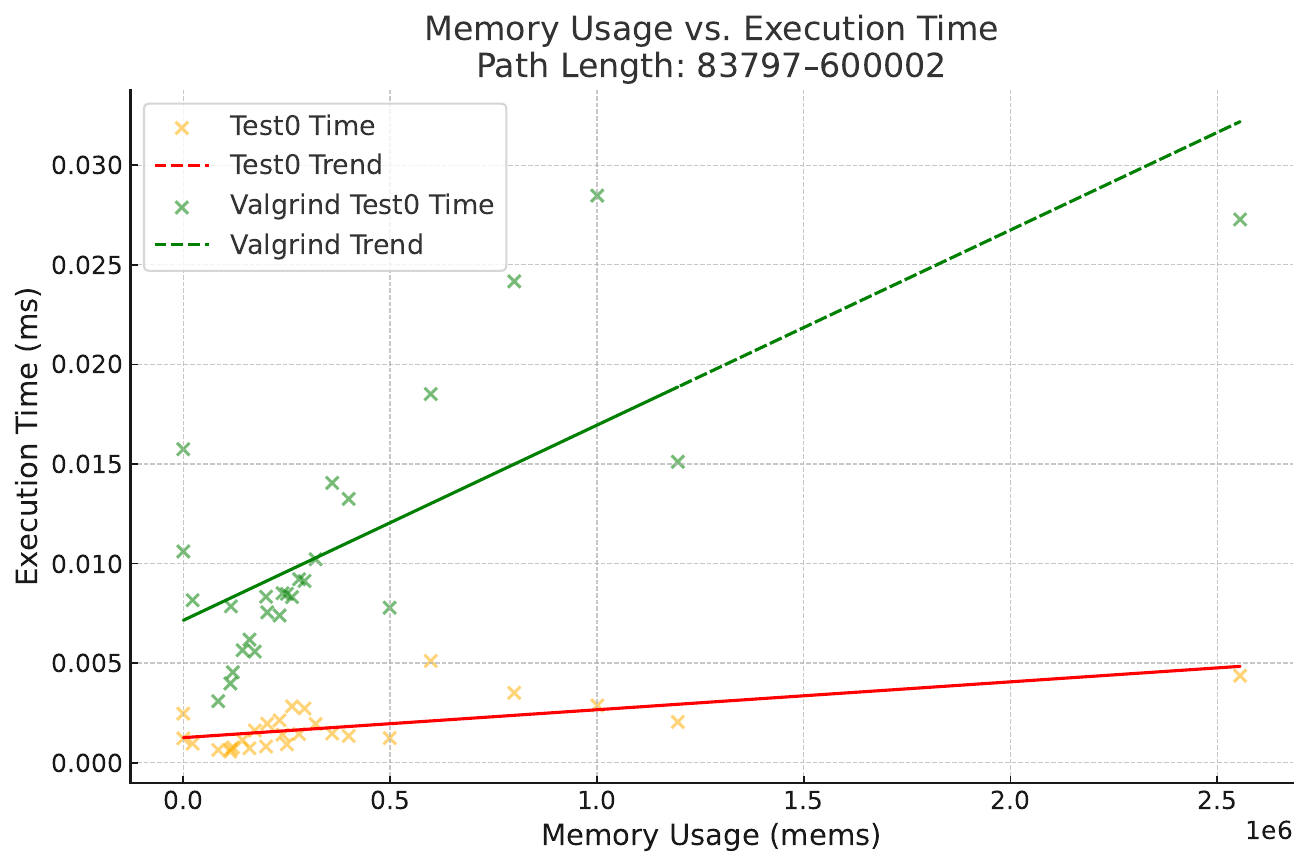}
    \includegraphics[width=0.48\textwidth]{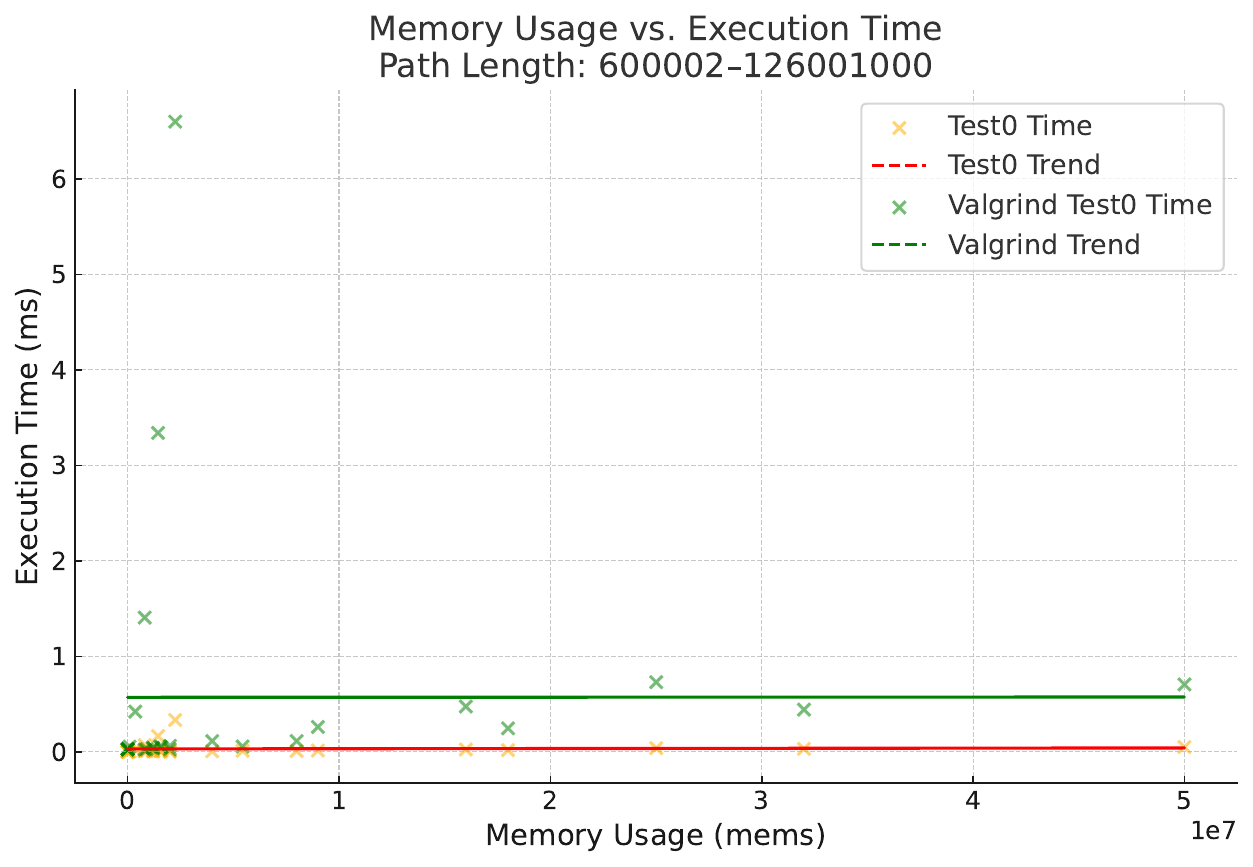}\\
    \caption{Scatter plots depicting the relationship between \texttt{mems} and execution time after grouping data by similar \texttt{path\_length}.}
    \label{fig:mems_exec_time_grouped_by_pathlength}
\end{figure*}

To further explore this issue, Fig.~\ref{fig:mems_exec_time_grouped_by_pathlength} divides the dataset into subsets grouped by similar \texttt{path length}, examining more closely the relationship between \texttt{mems} and execution time within each subgroup. Within these finer subdivisions, correlations vary significantly and sometimes deviate from the overall trend, emphasizing the complexity and variability of these relationships. These results highlight the need for more nuanced interpretation when using \texttt{mems} as a direct performance metric across diverse program paths.

\begin{table}[t]
\centering
\caption{Selected Experimental Results on Server}
\small
\begin{tabular}{l|c|r|c}
\hline
\textbf{Program} & \textbf{Path Length} & \textbf{Mems} & \textbf{Time (ms)} \\
\hline
\texttt{bubble.c (100)} & ~9999 & 19800 & 1.857 \\
\texttt{change.c (100)} & 10106 & 14 & 1.473 \\
\texttt{shell.c (1000)} & 12715 & 18800 & 2.743 \\
\texttt{sieve.c (5000)} & 13175 & 13089 & 1.267 \\
\texttt{array.c (5000)} & 15002 & 20000 & 1.479 \\
\texttt{FFT.c (2048)} & 18397 & 118592 & 7.134 \\
\hline
\end{tabular}
\label{table:server_results}
\end{table}

Table~\ref{table:server_results} provides a selection of representative results from six programs tested on the server. For instance, while both \texttt{bubble.c} and \texttt{shell.c} have similar \texttt{mems} values (19800 vs. 18800), the runtime of \texttt{shell.c} is noticeably longer, implying differences in loop structure or arithmetic cost. A more dramatic deviation is seen in \texttt{change.c}, which has a long path length but very low \texttt{mems} (only 14), yet its execution time is comparable to more memory-intensive programs. This confirms that some programs, especially those with simple arithmetic or control-heavy operations but minimal memory activity, can exhibit low \texttt{mems} but still consume substantial runtime.


Overall, these server-side experiments reveal that although \texttt{mems} is often correlated with execution time, its predictive ability is not uniform across all program types. Memory access count provides useful but incomplete insight into performance behavior. This reinforces the importance of combining \texttt{mems} with other indicators (e.g., arithmetic intensity or data dependencies) for more robust performance modeling.

\subsubsection{Local Machine Experiments: \noindent Stronger Trends Under Varied Memory Loads}
\noindent To account for the influence of high-performance hardware on timing variability, additional experiments were conducted on a local Windows 11 environment, disabling compiler optimizations (\texttt{-O0}) to minimize measurement noise. Execution times were measured using \texttt{QueryPerformanceCounter} instead of \texttt{valgrind}.

\begin{table}[t]
\centering
\small
\setlength{\tabcolsep}{4pt}
\caption{Selected Experimental Results on Local Machine}
\begin{tabular}{l|r|r|r}
\hline
\textbf{Program} & \textbf{Path Length} & \textbf{Mems~~} & \textbf{Time (ms)} \\
\hline
\texttt{bubble (500)} & 249999 & 748500 & 117.102 \\
\texttt{change (500)} & 250506 & 14 & 167.743 \\
\texttt{sieve (300000)} & 908472 & 907928 & 289.402 \\
\texttt{bubble (1000)} & 999999 & 2997000 & 485.858 \\
\texttt{insertsort (5000)} & 12507498 & 25004998 & 14913.156 \\
\texttt{bubble (4000)} & 15999999 & 47988000 & 28550.721 \\
\hline
\end{tabular}
\label{table:local_results}
\end{table}

Table~\ref{table:local_results} summarizes selected results from our local machine experiments, highlighting key findings related to the relationship between \texttt{mems} and execution time.
The experiments illustrate that while paths within the same or similar programs generally demonstrate predictable trends—such as paths with significantly higher memory access counts (\texttt{mems}) typically exhibiting longer execution times—this correlation does not universally hold across different programs. For instance, comparing the programs \texttt{bubble(500)} and \texttt{change(500)}, both having nearly identical path lengths, reveals a notable anomaly: despite \texttt{bubble} featuring substantially greater \texttt{mems} (748,500 vs 14), it executes faster (117.1ms vs 167.7ms). 

Such inconsistencies emphasize that while \texttt{mems} effectively differentiates between paths with stark contrasts in memory intensity, it does not reliably predict execution times across structurally diverse programs or different algorithmic classes. Factors such as cache locality, loop complexity, arithmetic operation intensity, and control-flow structure likely contribute to these deviations, underscoring the need for additional metrics or combined analysis strategies to achieve more accurate cross-program performance predictions.

\begin{figure}[htbp]
\centering
\includegraphics[width=\columnwidth]{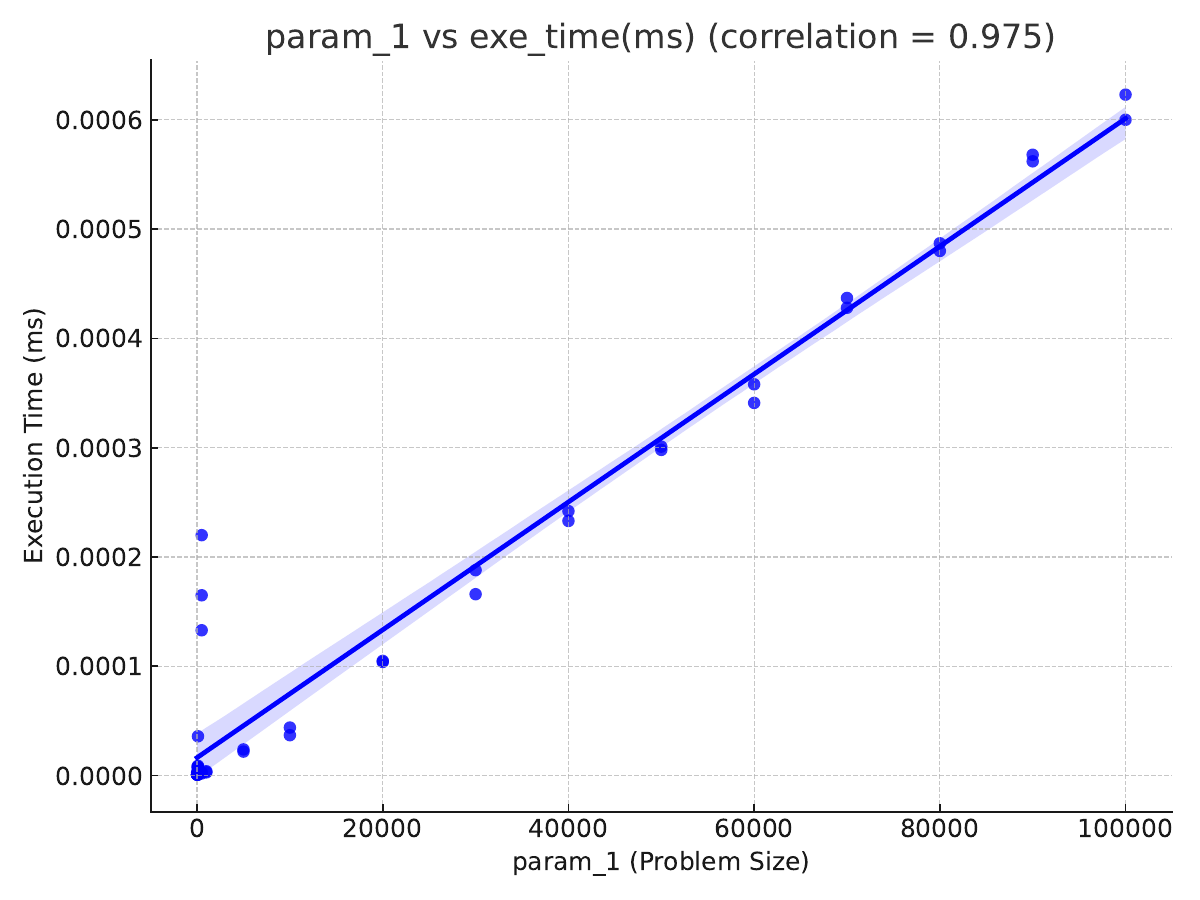}
\caption{Scatter Plot of Input Size vs. Execution Time}
\label{fig:scatter_n_time}
\end{figure}

The experimental outcomes reveal that while mems generally exhibit a positive correlation with execution time, this relationship is frequently inconsistent or obscured by other variables such as algorithmic characteristics and input size. Notably, input size  itself demonstrated a strong correlation (0.9736) with execution time, overshadowing mems as a standalone predictor. Figure \ref{fig:scatter_n_time} illustrates the clear positive trend between input size and execution time, emphasizing that larger input sizes consistently lead to longer execution times.

\subsubsection{Single-Core Execution: Limited Gains, Persistent Anomalies}

\noindent Considering the possibility that multi-core execution environments could introduce bias or inconsistencies into our experimental results, we restricted the experiments to single-core execution. In Windows, this single-core CPU affinity was enforced using the command: \texttt{cmd /c start /affinity 1}. We reused our previously established benchmarks and compared their results in single-core versus multi-core settings. 

\begin{figure*}[htbp]
    \centering
    \includegraphics[width=\linewidth]{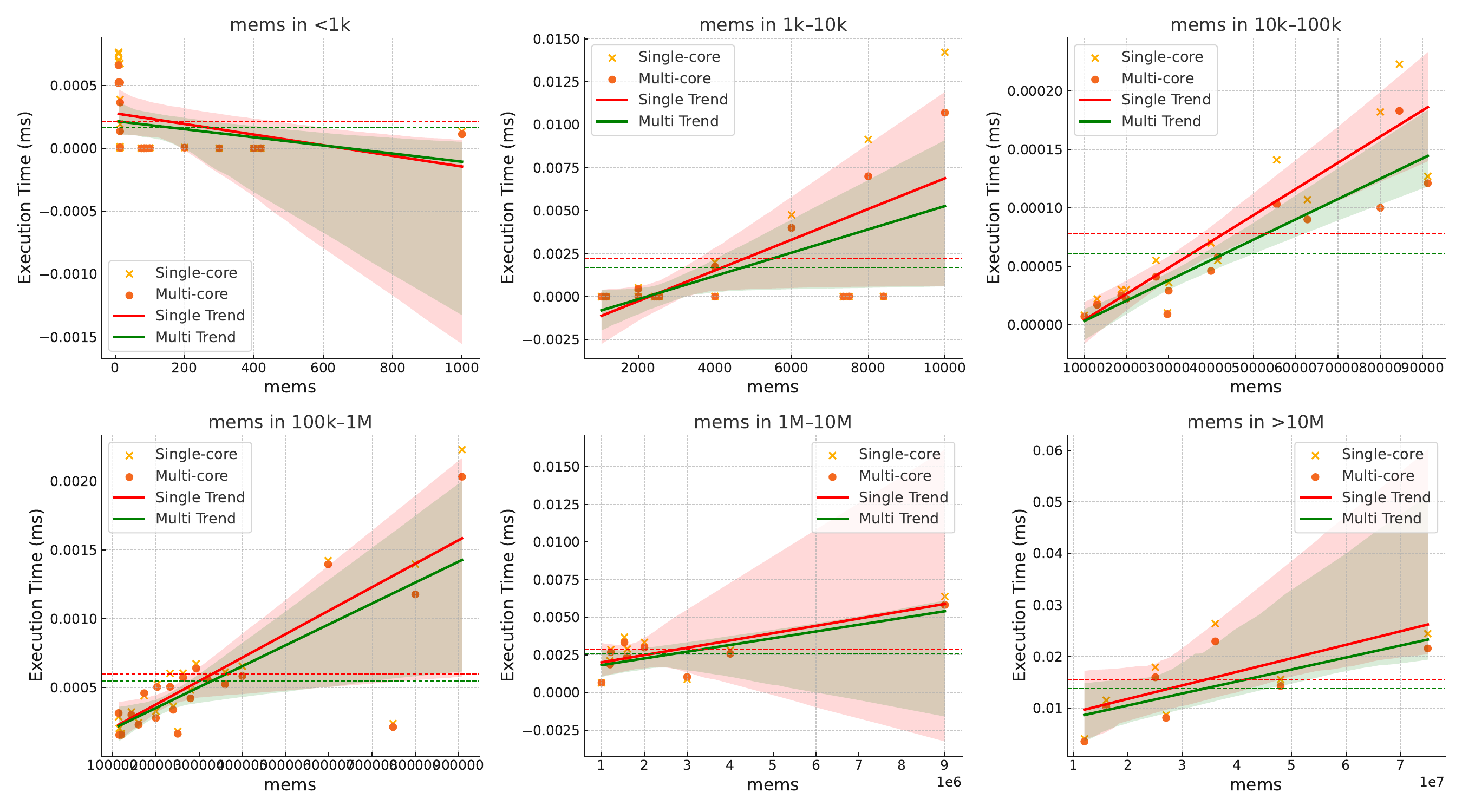}
    \caption{Execution time distributions comparing single-core and multi-core executions, grouped by memory accesses (mems) magnitude intervals.}
    \label{fig:cpu_mems}
\end{figure*}

To clearly understand how execution times vary under different memory access (mems) intensities, we partitioned the data into six magnitude-based intervals: $<1K$, $1K$–$10K$, $10K$–$100K$, $100K$–$1M$, $1M$–$10M$, and $>10M$. Figure~\ref{fig:cpu_mems} visually presents the execution time distributions for multi-core and single-core environments across these intervals.

\begin{table}[t]
\caption{Speedup Ratios Between Single-core and Multi-core Executions Across Different Mems Intervals}
\label{tab:speedup_ratios}
\centering
\resizebox{\columnwidth}{!}{
\begin{tabular}{c|c|c|c}
\hline
\textbf{Mems} & \textbf{Single-core Time (ms)} & \textbf{Multi-core Time (ms)} & \textbf{Speedup Ratio} \\  \hline
$<$1k        & 0.000217 & 0.000170 & 1.28$\times$ \\ 
1k–10k       & 0.002192 & 0.001709 & 1.28$\times$ \\ 
10k–100k     & 0.000078 & 0.000061 & 1.29$\times$ \\ 
100k–1M      & 0.000599 & 0.000550 & 1.09$\times$ \\ 
1M–10M       & 0.002842 & 0.002593 & 1.10$\times$ \\ \hline
\end{tabular}}
\end{table}

From Fig.~\ref{fig:cpu_mems}, it can be observed that multi-core execution consistently outperforms single-core execution across all intervals. Nonetheless, the performance gains differ notably among intervals. Specifically, the largest average speed-up ratio of approximately $1.29\times$ is observed within the intermediate load interval ($10^4$–$10^5$ mems). In contrast, while multi-core configurations maintain advantages at high load intervals (above $10^6$ mems), the speed-up ratios tend to stabilize and diminish slightly due to system-level limitations such as scheduling overheads and memory bandwidth bottlenecks.

Table~\ref{tab:speedup_ratios} provides the quantified speedup ratios when comparing execution times under single-core versus multi-core environments across different mems intervals. As previously noted, the multi-core configuration consistently outperforms the single-core environment, with peak acceleration of approximately 1.29$\times$ observed in the interval of $10^4$–$10^5$ mems. However, for larger intervals (above $10^5$ mems), the speedup becomes less pronounced, likely due to inherent system constraints such as memory bandwidth limitations and scheduling overhead.

Though restricting execution to a single-core environment indeed impacts certain benchmarks, the overall difference is relatively modest compared to amplification techniques such as Valgrind. Still, variations exist among different programs, suggesting nuanced effects rather than a uniform scaling factor. For instance, Table~\ref{tab:single_core_examples} provides detailed results comparing single-core and multi-core executions for selected representative test cases:

\begin{table}[t]
\caption{Single-Core Execution Examples}
\label{tab:single_core_examples}
\centering
\resizebox{\columnwidth}{!}{
\begin{tabular}{l|l|c|r|c}
\hline
\textbf{Program} & \textbf{Path File} & \textbf{Path Length} & \textbf{Mems~} & \textbf{Time (s)} \\ \hline
array & path\_13 & 240002 & 320000 & 0.134836 \\ 
bubble & path\_3 & 249999 & 748500 & 0.250803 \\ 
change & path\_2 & 250506 & 14 & 0.188000 \\ 
sieve & path\_14 & 263282 & 262985 & 0.200957 \\ \hline
\end{tabular}}
\end{table}

Similar to the multi-core experiments, comparisons involving the \texttt{change.c} program exhibit a clear positive correlation between memory access counts and execution time. Nevertheless, we observe notable exceptions. Specifically, the bubble sort program, despite exhibiting a significantly greater path length and higher mems than the array program, paradoxically displays less than half the execution time. Such anomalies suggest that neither single-core nor multi-core execution environments alone account fully for observed deviations. Hence, additional factors, possibly including data locality, memory caching patterns, algorithmic complexity, and even compiler optimizations, must be considered to explain these discrepancies adequately.

In summary, while single-core execution control provides additional clarity regarding the influence of CPU concurrency, it confirms that substantial variations in execution times and memory access correlations cannot be explained solely by CPU affinity. This indicates a necessity for exploring further system and algorithm-specific characteristics that influence runtime performance.

\textbf{Answer to RQ2 (Finding 2).} 
When comparing across different programs, \texttt{mems} exhibits partial correlation with execution time but lacks consistency as a standalone metric. While programs with significantly higher \texttt{mems} values often demonstrate longer runtimes, numerous counterexamples reveal that factors such as algorithmic structure, loop nesting, and data access patterns can dominate performance behavior. In many cases, execution time is more strongly correlated with input size or implementation-specific details rather than \texttt{mems} alone. Additionally, variations across platforms and compiler configurations further diminish the reliability of \texttt{mems} as a cross-program performance predictor. These findings suggest that \texttt{mems} can be informative but should be interpreted in conjunction with other structural metrics such as path length or input scale to yield reliable performance insights.

\section{Discussion}
\noindent Theoretical foundations, as discussed in Section II, suggest that memory access counts (\textit{mems}) are meaningful indicators of execution time. Both Knuth's foundational work~\cite{Knuth1994} and later efforts such as~\cite{Zhang2013} advocate for the use of memory operations as a proxy for computational cost. From a hardware perspective, memory accesses—especially to main memory—incur significantly higher latency than cache accesses or arithmetic operations, sometimes by orders of magnitude~\cite{hennessy2011computer}. This expectation is supported by our RQ1 experiments: within the same program, execution paths with more memory operations consistently showed longer execution times, reinforcing the theoretical assumptions.

However, RQ2 highlights limitations in generalizing this trend across different programs. Even when controlling for path length, we observed notable inconsistencies between mems and actual execution time. These deviations are likely rooted in program-level structural and algorithmic differences. For example, two programs with similar control-flow depth might exhibit different memory access behaviors depending on their internal loop patterns, data locality, or access regularity. 
A path with higher mems may still execute faster due to better cache performance or vectorization-friendly loops.

Another contributing factor is input scale. We found that input size ($n$) itself showed a strong correlation with execution time, often overshadowing the role of mems. This suggests that in multi-program comparisons, input characteristics and algorithmic design may dominate the performance profile, making mems insufficient as a standalone metric.

To isolate the effect of mems more effectively, future work may benefit from stricter experimental controls—such as comparing programs with similar memory access patterns, input shapes, and loop structures. While such controls can improve causal attribution, they also reduce benchmark diversity and may limit the generalizability of results.

Hardware-specific factors also played a role. Our experiments involved multiple environments to capture variations in CPU architecture and compiler optimizations. While we controlled for optimization flags (e.g., using \texttt{-O0}), we did not explicitly control cache behavior. Interestingly, we found little difference in mems–time correlation between single-core and multi-core configurations, suggesting that deeper architectural factors (e.g., cache policies or prefetching) may be more influential than core count.

Our study also uncovered secondary findings. Notably, instrumentation overhead—particularly from \texttt{printf} calls—added measurable latency. This overhead sometimes masked the true runtime differences attributable to mems, especially in short-running programs. For more precise measurement in future work, such overhead should be minimized, for instance by using buffered logging or low-overhead profiling mechanisms.

Finally, we acknowledge limitations in our benchmark suite. The programs we tested were algorithmically representative but limited in scale and domain diversity. Most benchmarks involved short absolute runtimes, which may have introduced timing noise and biased results. Nonetheless, repeated experiments under varied conditions revealed consistent overall trends, suggesting that the conclusions are robust. Future work should expand on this foundation by incorporating more real-world applications, longer-running programs, and broader architectural platforms to fully assess the reliability of \textit{mems} as a static performance estimator.

\section{Related Work}

\noindent Performance testing and analysis is a long-standing topic in software engineering. A representative early work is WISE~\cite{Burnim2009} later enhanced this idea with a hybrid approach. This idea was later extended by Noller et al.~\cite{Noller2018} with a hybrid approach,combining fuzzing and symbolic execution for improved coverage in Java programs.

However, subsequent studies found that such static indicators often fail to capture real runtime behavior~\cite{Schnoor2020}. To improve accuracy, recent work incorporates dynamic profiling or hybrid metrics that combine static features with empirical timing. These methods provide better precision but increase overhead and reduce portability.

In real-time and embedded systems, Worst-Case Execution Time (WCET) remains a key metric~\cite{Wilhelm2008,Bundala2015}, estimating the upper bound of execution time for safety-critical applications. WCET tools are typically hardware-specific, limiting their generalizability. Since WCET and WISE both aim to characterize worst-case performance, using architecture-independent static performance metrics—like memory access counts (\textit{mems})—offers a potential middle ground: enabling static estimation of performance costs while remaining generalizable across platforms.

Our work follows this direction by validating \textit{mems} as a lightweight, path-sensitive static metric that may support WCET-style reasoning without relying on platform-specific tools. For broader test efficiency, Cheng et al.~\cite{Cheng2021} show how performance-aware prioritization can improve configuration testing, highlighting the value of performance metrics even outside traditional timing analysis.

\section{Conclusion}

\noindent 
The metric, \textit{mems}, can be used to statically calculate a program's performance.
In this paper, we provide automated solutions based on the new metric and evaluate it in real programs. 
We have found that in the same program, this metric has a significant positive correlation with the program execution time; while in different programs, the correlation becomes less consistent due to significant differences in memory access patterns and internal structural complexity.
We hope that this metric and such findings can help the research community. 
In the future, we will further investigate the applicability of this metric and extend our analysis to larger and more diverse codebases to strengthen its impact on software quality assurance.

\balance


\begin{thebibliography}{1}

\bibitem{Zhang2013}
J. ~Zhang, ``Performance estimation using symbolic data,'' in \textit{Theories of Programming and Formal Methods}, Lecture Notes in Computer Science, vol.~8051, pp.~346--353, Springer, 2013.

\bibitem{Boyer1977}
R.~Boyer and J.~Moore, ``A fast string matching algorithm,'' \textit{Communications of the ACM}, vol.~20, pp.~762--772, 1977.

\bibitem{Cole1991}
R.~Cole, ``Tight bounds on the complexity of the Boyer-Moore string matching algorithm,'' in \textit{Proc. of the 2nd ACM-SIAM Symposium on Discrete Algorithms (SODA)}, pp.~224--233, 1991.

\bibitem{Geldenhuys2012}
J.~Geldenhuys, M.~B.~Dwyer, and W.~Visser, ``Probabilistic symbolic execution,'' in \textit{Proc. of the International Symposium on Software Testing and Analysis (ISSTA)}, pp.~166--176, 2012.

\bibitem{King1976}
J.~C.~King, ``Symbolic execution and program testing,'' \textit{Communications of the ACM}, vol.~19, no.~7, pp.~385--394, 1976.

\bibitem{Knuth1994}
D.~E.~Knuth, \textit{The Stanford GraphBase: A Platform for Combinatorial Computing}. ACM Press, 1994. 

\bibitem{Liu2011}
S.~Liu and J.~Zhang, ``Program analysis: From qualitative analysis to quantitative analysis,'' in \textit{Proc. of the 33rd International Conference on Software Engineering (ICSE)}, pp.~956--959, 2011.

\bibitem{Ma2009}
F.~Ma, S.~Liu, and J.~Zhang, ``Volume computation for Boolean combination of linear arithmetic constraints,'' in R.~A.~Schmidt, Ed., \textit{Proc. of CADE-22}, \textit{LNCS}, vol.~5663, Springer, pp.~453--468, 2009.

\bibitem{Zhang2004}
J.~Zhang, ``Quantitative analysis of symbolic execution,'' presented at the \textit{28th International Computer Software and Applications Conference (COMPSAC)}, 2004.

\bibitem{Zhang2008}
J.~Zhang, ``Constraint solving and symbolic execution,'' in B.~Meyer and J.~Woodcock, Eds., \textit{Proc. of VSTTE 2005}, \textit{LNCS}, vol.~4171, Springer, pp.~539--544, 2008.

\bibitem{Zhang2013draft}
J.~Zhang, S.~Liu, and F.~Ma, ``A tool for computing the volume of the solution space of SMT (LAC) constraints,'' unpublished draft, Jan. 2013.

\bibitem{Burnim2009}
J.~Burnim, S.~Juvekar, and K.~Sen, ``WISE: Automated test generation for worst-case complexity,'' in \textit{Proc. of the 31st International Conference on Software Engineering (ICSE)}, Vancouver, Canada, pp.~463--473, May 2009.

\bibitem{Noller2018}
Y.~Noller, R.~Kersten, and C.~S.~Pasareanu, ``Badger: Complexity analysis with fuzzing and symbolic execution,'' in \textit{Proc. of the 27th ACM SIGSOFT International Symposium on Software Testing and Analysis (ISSTA)}, Amsterdam, The Netherlands, pp.~322--332, July 2018.

\bibitem{Wilhelm2008}
R.~Wilhelm, J.~Engblom, A.~Ermedahl, N.~Holsti, S.~Thesing, D.~Whalley, G.~Bernat, C.~Ferdinand, R.~Heckmann, T.~Mitra, \textit{et al.}, ``The worst-case execution-time problem—overview of methods and survey of tools,'' \textit{ACM Transactions on Embedded Computing Systems (TECS)}, vol.~7, no.~3, pp.~1--53, 2008.

\bibitem{Schnoor2020}
H.~Schnoor and W.~Hasselbring, ``Comparing static and dynamic weighted software coupling metrics,'' \textit{Computers}, vol.~9, no.~2, p.~24, 2020.

\bibitem{Knuth1997}
D.~E.~Knuth, \textit{The Art of Computer Programming}, 2nd ed., vols.~1--3; 1st ed., vol.~4A, Addison-Wesley, Reading, Massachusetts, 1997. Fascicles of Volume~4 in progress.

\bibitem{Saumya2019}
C.~Saumya, J.~Koo, M.~Kulkarni, and S.~Bagchi, ``XSTRESSOR: Automatic generation of large-scale worst-case test inputs by inferring path conditions,'' in \textit{Proc. of the 12th IEEE International Conference on Software Testing, Validation and Verification (ICST)}, pp.~1--12, IEEE, 2019.

\bibitem{Bundala2015}
D.~Bundala and S.~A.~Seshia, ``On systematic testing for execution-time analysis,'' \textit{CoRR}, vol.~abs/1506.05893, 2015. [Online]. Available: \url{http://arxiv.org/abs/1506.05893}

\bibitem{Chen2022}
W.~Chen, C.~Tatsuoka, and X.~Lu, ``HiBGT: High-performance Bayesian group testing for COVID-19,'' in \textit{Proc. of the 29th IEEE International Conference on High Performance Computing, Data, and Analytics (HiPC)}, pp.~176--185, IEEE, 2022.

\bibitem{Ma2009}
F.~Ma, S.~Liu, and J.~Zhang, ``Volume computation for boolean combination of linear arithmetic constraints,'' in \textit{Proc. of the 22nd International Conference on Automated Deduction (CADE)}, ser. Lecture Notes in Computer Science, vol.~5663, pp.~453--468, Springer, 2009.

\bibitem{Ge2021}
C.~Ge and A.~Biere, ``Decomposition strategies to count integer solutions over linear constraints,'' in \textit{Proc. of the 30th Int'l Joint Conf. on Artificial Intelligence (IJCAI)}, pp.~1389--1395, 2021.

\bibitem{eppather}
L.~Zhang, ``\textit{eppather: A Test Data Generation Tool for Unit Testing of C Programs},'' GitHub Repository, 2024. Available: \url{https://github.com/Z769018860/eppather}.

\bibitem{hennessy2011computer}
J.~L.~Hennessy and D.~A.~Patterson, \textit{Computer Architecture: A Quantitative Approach}, 5th~ed. Morgan Kaufmann, 2011.

\bibitem{Cheng2021}
R.~Cheng, L.~Zhang, D.~Marinov, and T.~Xu, ``Test-case prioritization for configuration testing,'' in \textit{Proc. of the 30th ACM SIGSOFT Int'l Symp. on Software Testing and Analysis (ISSTA)}, pp.~452--465, ACM, 2021.

\bibitem{XuZhang2006}
Z.~Xu and J.~Zhang, ``A test data generation tool for unit testing of C programs,'' in \textit{Proceedings of the 6th International Conference on Quality Software (QSIC)}, Beijing, China, Oct. 2006, pp.~107--116.

\bibitem{JZhang2004}
J.~Zhang, ``Symbolic execution of program paths involving pointer and structure variables,'' in \textit{Proceedings of the 4th International Conference on Quality Software (QSIC)}, Braunschweig, Germany, Sep. 2004, pp.~87--92.

\bibitem{Knuth2012}
D.~E.~Knuth, ``Satisfiability and The Art of Computer Programming,'' in \textit{Proc. of the 15th Int'l Conf. on Theory and Applications of Satisfiability Testing (SAT)}, Trento, Italy, Jun. 2012, vol.~7317, Lecture Notes in Computer Science, p.~15, Springer.

\bibitem{Chen2016}
B.~Chen, Y.~Liu, and W.~Le, ``Generating performance distributions via probabilistic symbolic execution,'' in \textit{Proc. of the 38th Int'l Conf. on Software Engineering (ICSE)}, pp.~49--60, ACM, 2016.





\end{thebibliography}
\end{document}